\documentclass[referee,useAMS,usenatbib]{mn2e}
\usepackage{graphicx}
\title[INOV of PKS0735+178]
{Unusual optical quiescence of the classical BL Lac object PKS 0735+178
on intranight time scale }
\author[Goyal et al.]
  {Arti Goyal$^1$\thanks{E-mail: arti@aries.res.in},
   Gopal-Krishna$^{2}$,
   G. C. Anupama$^{3}$,
   D. K. Sahu$^{3}$,
   R. Sagar$^{1}$,
\newauthor
   S. Britzen$^4$,
   M. Karouzos$^{4}$\thanks{Member of the International Max Planck Research School (IMPRS) for Astronomy and Astrophysics at the Universities of Bonn and Cologne},
   M. F. Aller$^{5}$, 
   H. D. Aller$^{5}$\\ 
$^1$ Aryabhatta Research Institute of observational sciencES (ARIES),
Manora Peak, Naini Tal 263 129, India\\
$^2$ NCRA.TIFR, Pune University Campus, Pune 411 007, India\\ 
$^3$ Indian Institute of Astrophysics (IIA) Bangalore 560 034, India \\ 
$^4$ Max-Planck-Institut f\"ur Radioastronomie, Auf dem H\"ugel 69, 53121 Bonn, 
Germany \\
$^5$ Astronomy Department, University of Michigan, Dennison Building, Ann Arbor, MI 48109-1090, USA \\
}
\date{Released 2008 Xxxxx XX}

\pagerange{\pageref{firstpage}--\pageref{lastpage}} \pubyear{2008}

\def\LaTeX{L\kern-.36em\raise.3ex\hbox{a}\kern-.15em
    T\kern-.1667em\lower.7ex\hbox{E}\kern-.125emX}

\begin{document}

\label{firstpage}

\maketitle

\begin{abstract}

We present the result of our extensive intranight optical monitoring 
of the well known low-energy peaked BL Lac (LBL) object PKS 0735+178. 
This long-term follow-up consists of $R$-band monitoring for a minimum 
duration of $\sim 4$ hours, on 17 nights spanning 11 years 
(1998-2008). Using the CCD as an N-star photometer, a detection limit 
of around $1\%$ was attained for the intra-night optical variability 
(INOV). Remarkably, an INOV 
amplitude of $\geq 3\%$ on hour-like time scale was not observed on
any of the 17 nights, even though the likelihood of a typical LBL 
showing such INOV levels in a single session of $\ga 4$ hours 
duration is known to be high ($\sim50\%$). Our observations have thus
established a peculiar long-term INOV quiescence of this radio-selected 
BL Lac object. Moreover, the access to unpublished optical monitoring 
data of similarly high sensitivity, acquired in another programme, 
has allowed us to confirm the same anomalous INOV quiescence of this 
LBL all the way back to 1989, the epoch of its historically largest 
radio outburst.  Here, we present observational evidence revealing 
the very unusual INOV behaviour of this classical BL Lac object and
discuss this briefly in the context of its other known exceptional
properties.

\end{abstract}

\begin{keywords}
BL Lac objects -- general - individual : PKS 0735+178 --
 variability -- intranight -- optical: Galaxies - active -- Galaxies - jets
\end{keywords}

\section{Introduction}

Characterized by extreme faintness or absence of broad emission lines in
their optical/UV spectra, BL Lac objects are a subset of the {\it blazar}
population for which the dominant source of emission is believed to be a
relativistic jet of non-thermal radiation 
(e.g., Blandford \& Rees 1978; Urry \& Padovani 1995). A key optical
property of blazars is large and frequently occuring rapid flux variability 
(also termed as INOV), which is not exhibited by other 
radio-loud or radio-quiet active galactic nuclei (AGN). Thus, a classical, 
radio-selected BL lac object (RBL), when monitored continuously for more 
than about 4 hours, is quite likely to show optical variability
at the level of a few percent, on hour-like time scale (e.g., Miller, Carini \& Goodrich 
1989; Wagner \& Witzel 1995; Romero et al. 2002). More specifically, INOV 
amplitude $\psi > 3\%$ is expected to occur in such observations, with a 
probability (duty cycle: DC) of $\sim 53\%$ (Gopal-Krishna et al. 2003; Stalin 
et al. 2005). A similar estimate (DC $\sim 57\%$) is obtained using the data 
from an independent programme of intranight monitoring of a well defined 
sample of EGRET detected RBLs (see Table 1 of Romero et al. 2002). For X-ray 
selected blazars (XBLs), the INOV duty cycle is about two times smaller
(Romero et al. 2002; also, Heidt et al. 1998). Recall that the above estimates 
for RBLs are based on light curves of durations longer than 4 hours and
therefore, for consistency, we shall stick to the same criterion while 
selecting any intranight light curves from the literature for the purpose 
of making statistical comparisons.

Since recent studies (Stalin et al. 2005; Gopal-Krishna et al. 2003) have 
shown that large amplitude INOV ($\psi > 3\%$) is displayed exclusively by 
blazars, i.e., BL Lacs and Highly Polarized Quasars (HPQs), such INOV is 
generally expected to be primarily associated with the relativistic jet. Yet,
the precise mechanism of the INOV remains obscure. The possibilities often 
invoked include perturbations in the inner jet, caused by shear or relativistic 
shocks, small-scale inhomogeneities in the magnetic field and instabilities in 
the particle acceleration, etc. (e.g., Wiita 1996; Marscher 1996 and 
references therein).
Some contribution to INOV can also be expected from accretion disk 
instabilities (Mangalam \& Wiita 1993) and, occasionally, from ``superluminal 
gravitational microlensing'' (Gopal-Krishna \& Subramanian 1991; also, Rabbette 
et al. 1996; Romero, Surpi \& Vucetich. 1995). Potential clues for assessing the
relative importance of these various mechanisms may emerge by identifying 
and then studying in detail 
blazars exhibiting some highly unusual pattern of INOV. 

During the ARIES programme of optical monitoring of a sample of powerful AGN, 
the well known RBL, PKS 0735+178, was found to exhibit remarkably low-level 
INOV ($\psi <\sim 1\%)$ on all the 4 nights it was monitored during 
1998-2001 (Sagar et al. 2004). 
To probe this unexpected result, we have extended intranight 
monitoring of this RBL for a further seven years. This has yielded its 
intranight $R$-band differential light curves (DLCs) for a total of 17 nights, 
between 1998 and 2008. Thus, the DLCs are available for minimum one night 
every winter since 1998, except for 2002. Each DLC is longer than $\sim 4$ 
hours and has a sensitivity adequate for a secure detection of INOV down 
to $1 - 2\%$ level. In this paper, we report these new data and discuss our 
entire dataset in the broader context of the published multi-wavelength 
observations of this well known BL Lac object.

PKS 0735+178 is among the first sources to be 
designated as ``classical BL Lac'' (Carswell et al. 1974). Its host galaxy 
still remains unresolved 
(e.g., Pursimo et al. 2002) but an absorption feature in the optical spectrum, 
identified as Mg-II, has yielded a lower redshift limit of $z > 0.424$ (e.g., 
Rector \& Stocke 2001).  Hartman et al. (1999) have reported $\gamma-$ray 
detection of this blazar using EGRET. On the basis of its spectral energy 
distribution peaking at $10^{13-14}$ Hz, it can be confidently classified as a 
`low energy peaked' BL Lac (LBL) (Padovani et al. 2006; Nieppola, Tornikoski 
\& Valtaoja 2006; Ghisellini, Tavecchio \& Chiaberge 2005). 
Whereas in the optical and even in the radio band PKS 0735+178 has shown 
strong variability (Webb et al. 1988; Ciprini et al. 2007; Qian \& Tao 2004;
Fig. 1; Sect. 4) characteristic of blazars, its X-ray and 
$\gamma-$ray emission was found to be quite steady (Bregman et al. 1984; 
Madejski \& Schwarz 1988; Nolan et al. 2003). Specifically, based on their 
analysis of the Einstein Observatory (IPC) observations from April, 1979 till 
March, 1981, Madejski \& Schwarz (1988) found no deviation from the mean 
flux density at $\sim$ 1 keV, by more than 20\%, on time scales between 
several hours to 2 years. Thus, they concluded that ``the lack of variability 
in PKS 0735+178 is exceptional among BL Lac objects'' (Sect. IIb of their 
paper). In contrast, its optical flux more than doubled during the same 
period (Fig. 1b). Another curious early finding was its remarkably 
smooth flat radio spectrum, which was interpreted as the superposition of many 
homogeneous components of incoherent synchrotron radiation, so called ``cosmic 
conspiracy'' (Cotton et al. 1980).

\section{Observations and data reduction}

The photometric observations were carried out using the 104-cm Sampurnanand 
telescope (ST) located at Aryabhatta Research Institute of observational 
sciencES (ARIES), Naini Tal (India), except on one night when the 201-cm 
Himalayan Chandra telescope (HCT) of IAO at Hanle (India) was used. ST has a 
Ritchey-Chretien (RC) optics with a f$/$13 beam (Sagar 1999). The detector 
was a cryogenically cooled 2048 $\times$ 2048 chip mounted at the Cassegrain 
focus. This chip has a readout noise of 5.3 e$^{-}$/pixel and a gain of 
10 e$^{-}$$/$Analog to Digital Unit (ADU) in an usually employed slow 
readout mode. Each pixel has a dimension of 24 $\mu$m$^{2}$ which corresponds
to 0.37 arcsec$^{2}$ on the sky, covering a total field of 13$^{\prime}$ 
$\times$ 13$^{\prime}$. Observations were carried out in 2 $\times$ 2 binned 
mode to improve the S$/$N ratio. The seeing mostly ranged between 
$\sim$1.5$^{\prime\prime}$ to $\sim$3$^{\prime\prime}$, as determined using 
three moderately bright stars recorded along with the blazar on the same CCD 
frame (Fig. 2). 

The HCT is located at the Indian Astronomical Observatory (IAO), Hanle (Ladakh)
in northern India. It is also of the RC design with a f$/$9 beam at the 
Cassegrain focus\footnote{http://www.iiap.res.in/$\sim$iao}. The detector was 
a cryogenically cooled 2048 $\times$ 4096 chip, of which the central 2048 
$\times$ 2048 pixels were used. The pixel size is 15 $\mu$m$^{2}$ so that 
the image scale of 0.29 arcsec$/$pixel covers an area of 10$^{\prime}$ 
$\times$ 10${^\prime}$ on the sky. The readout noise of CCD is 
4.87 e$^{-}$/pixel and the gain is 1.22 e$^{-}$$/$ADU. The CCD was used in 
an unbinned mode. 

All the observations were carried out using {\it R} filter for which the CCDs
used have maximum sensitivity. The exposure time was typically 12-30 minutes 
for the ST and about 3 minutes for the HCT observations. The 
field positioning was adjusted so as to also include within the CCD frame at
least 2-3 comparison stars within about a magnitude of the blazar, in order to 
minimize the possibility of spurious variability detection (see, e.g.,
Cellone, Romero \& Araudo 2007). For both the telescopes, the bias frames were 
taken intermittently and twilight sky flats were obtained. Table 1 gives
the log of our observations of this blazar, including those already reported 
in Sagar et al. (2004). The data provided include for each observation the 
date, the telescope used, number of data points in the DLC, the total 
duration of monitoring, the quantifiers of INOV, C$_{eff}$ and the amplitude 
$\psi$ (Sect. 3) as well as a remark on the INOV status.

The preprocessing of the CCD images (i.e. bias subtraction, flat-fielding and 
cosmic-ray removal) was done following the standard procedures in 
IRAF\footnote{\textsc {Image Reduction and Analysis Facility}} and 
MIDAS\footnote{\textsc {Munich Image and Data Analysis System}} packages. 
The instrumental magnitudes of the blazar and the stars in the image frames 
were determined by aperture photometry, using 
DAOPHOT \textrm{II}\footnote{\textsc {Dominion Astrophysical Observatory 
Photometry} software} (Stetson 1987). The magnitudes of the blazar were measured 
relative to the nearby comparison stars present on the same CCD frame in 
order to account for the extinction of blazar's light due to the earth's 
atmosphere. This way, Differential Light Curves (DLCs) of the blazar were 
generated relative to three comparison stars. Likewise, DLCs were also 
generated for each comparison star relative to the other two comparison stars. 
For each night, the selection 
of the optimum aperture radius for photometry was done on the basis of the 
observed dispersions in the star-star DLCs generated using different aperture 
radii, starting from the median seeing (FWHM) value on that night to 4 times 
that value. The aperture selected was the one which showed minimum scatter for 
the steadiest DLC found for the various pairs of the comparison stars. The
selected aperture radius was then used to generate DLCs for the target blazar 
relative to the comparison stars, as well as for the individual comparison 
stars, by pairing them with the remaining two comparison stars. The `seeing' 
was monitored throughout the night using three moderately bright stars 
recorded in each CCD frame. Additional details of the data reduction 
procedure can be found in Stalin et al. (2004, 2005). 

\section{Results}

Figure 2 shows our newly obtained intranight DLCs for PKS 0735+178 and the 
comparison stars, along with the plots of `seeing', as described above. By 
combining these with the DLCs for the four nights, published in Sagar et al. 
(2004), we have generated `long-term (differential) light curve', relative 
to the {\it same} set of comparison stars as used by Sagar et al. (2004), thus
maintaining continuity in the long-term optical DLC (Fig. 3). 
{\bf The optical field of PKS 0735+178, marking the blazar and the comparison stars 
used by us, is shown in Figure 4.}
The long-term 
DLC exhibits a peak-to-peak variation of $\sim$1.7 mag (Fig. 3).  We have 
also transformed these differential light curves into actual light curves, 
by computing the average instrumental magnitude for each of those nights and 
calibrating those values using at least two standard stars available on the 
CCD frames, after ensuring that they had remained unsaturated throughout the 
monitoring. The standard stars used are: C7 and D (as given by Ciprini et al. 
2007). The calibrated LTOV data points are plotted in Fig. 1a.

Based on the intranight DLCs (Fig. 2), we have determined the INOV status 
and the INOV amplitude ($\psi$) for each night, as given in Table 1. 
Table 2 lists the positions and apparent magnitudes of the comparison stars 
used in producing the INOV and LTOV differential light curves (Figs. 2 \& 3). 
The INOV classification `variable' (V) or `non-variable' (N) was decided using a 
computed parameter C$_{eff}$, basically following the criteria of Jang \& Miller 
(1997). We define C for a given DLC as the ratio of its standard deviation, 
$\sigma$$_T$ and $\eta\sigma_{err}$, where $\sigma_{err}$ is the average of the 
rms errors of its individual data points and $\eta$ was estimated to be 1.5 
(Stalin et al. 2004, 2005; Gopal-Krishna et al. 2003; Sagar et al. 2004).
However, our analysis for the present dataset yields $\eta = 1.3$ which we 
have used here.  We finally computed C$_{eff}$ for a given observing session, 
using the C values (as defined above) determined for the DLCs of the blazar
relative to different comparison stars monitored during that session (for
details, see Sagar et al. 2004). This procedure has the advantage of utilizing
the multiple DLCs of an AGN available during a single session (i.e., relative to different comparison 
stars). The source is termed `V' if C$_{eff}$$ > $ 2.57, corresponding to a 
confidence level of $99\%$. We call the AGN to be `probable variable' (PV) 
if C$_{eff}$ is found to be in range of 1.95 to 2.57, corresponding to a 
confidence level between $95\%$ to $99\%$. Finally, the peak-to-peak INOV 
amplitude ($\psi$) was calculated using the definition (Romero, Cellone \& Combi 1999):

\begin{equation}
\psi= \sqrt{({D_{max}}-{D_{min}})^2-2\sigma^2}
\end{equation}
with \\
$D_{max}$ = maximum in the AGN differential light curve \\
$D_{min}$ = minimum in the AGN differential light curve \\
$\sigma^2$= $\eta^2$$\langle\sigma^2_{err}\rangle$
\\

The INOV duty cycle (DC) for PKS 0735+178 was computed using our entire dataset 
of 17 nights, following the definition of Romero, Cellone \& Combi (1999)
(see, also, Stalin et al. 2004): 

\begin{equation}
DC  = 100\frac{\sum_{i=1}^n N_i(1/\Delta t_i)}{\sum_{i=1}^n (1/\Delta t_i)}\%
\label{eqno1}
\end{equation}
where $\Delta t_i = \Delta t_{i,obs}(1+z)^{-1}$ is the
duration of monitoring session of a source on the $i$th night, corrected for
the blazar's cosmological redshift $z$. Note that since the source was not
monitored for identical duration on each night, the computation has been weighted
by the actual duration of monitoring $\Delta t_i$. $N_i$ was set equal to 1 if 
INOV was detected, otherwise $N_i$ = 0.

DC is found to be $\sim 27 \%$, which increases to $ \sim 42 \% $ if the 
two cases of probable INOV are also included. The key result, however, is that 
{\it large INOV ($\psi \geq 3\%$) was consistently absent on all the 
17 nights}, even though the data quality remained adequate throughout.

\section{Summary of the variability patterns}

In this section we summarize the complex flux variability patterns exhibited 
by this blazar, in order to focus attention on both its normal and anomalous aspects.
Such a background perspective is important for appreciating its rather
surprising INOV behavior established in this work.

\subsection{Long-term optical variability}

Recently, Ciprini et al. (2007) have published a long-term $B$-band light curve 
of PKS 0735+178, spanning almost 100 years (1906 - 2004) (see, also Fan et al. 
1997). Of this, the best sampled segment covers the last 33 years 
(867 nights, 1970 onwards). In Fig. 1b we reproduce the light curve\footnote{
Data points were retrieved from Fig. 4 [lower panel] of Ciprini et al. (2007)
 using the standalone version of a programme {\it Dexter} 
(http://vo.uni-hd.de/dexter/ui/ui/custom) available over SAO/NASA ADS
by Demleitner et al. (2001).}, taking median of the data binned into 
successive one-year intervals.
Starting from Feb., 1993, the data given in Ciprini et al. (2007) are based on 
CCD monitoring ($BVRI$), with the densest sampling attained in the $R$-band. 
Again, we have taken the medians for successive 1-year bins and augmented those
data with our own $R$-band measurements for the period 1998-2008, after converting 
the calibrated magnitudes to flux densities and averaging over each night 
(Sect. 3).  The composite $R$-band light curve for the period 1993-2008 is shown 
in Fig. 1a.  It is found that the blazar's optical flux dropped close to the
historical minimum (occuring in early 1997), at the epoch 2007.0 from  
where it doubled by the end of 2007 and then dropped back in 2008 to a level 
close to the historical minimum. Considerably more pronounced
optical variability was recorded during the previous 7 years (Fig. 1a). Fig. 2 
of Ciprini et al. (2007) shows typical variations of about 1 mag on time 
intervals smaller than 1/2 year (see below), which together with the dominant 
radio core (see Sect. 1) is consistent with its being a LBL type BL Lac 
(e.g., Lister \& Smith 2000).

The 100-year optical light curve of PKS 0735+178 shows five optical outbursts, 
with the historical maximum (B $\sim$ 13.9 mag) reached in mid-1977 (Ciprini 
et al. 2007). The last major outburst occurred around mid-2001 
(B $\sim$ 15.0 mag), following the historical minimum (B $\sim$17.5 mag) 
in early 1997. Ciprini et al. (2007) have identified three main temporal components 
in the light curve: $\sim$ 4.5 yr, $\sim$ 8.5 yr and $\sim$ 12 yr. 
The shortest of these characteristic time scales had earlier been noted by 
Webb et al. (1988) and by Smith \& Nair (1995), whereas the intermediate 
time scale was inferred previously by Qian \& Tao (2004). Ciprini et al.
(2007) have suggested that these three characteristic 
time scales may well be harmonic signatures of one fundamental component of 
about 4 years. Note that the optical spectral index ($\alpha_{o}$; defined as 
$S_{\nu}\propto{\nu^{-\alpha}}$) given by Ciprini et al. (2007) for the data set 
1993-2004 did not show any correlation with the fluctuations in the light 
curve (upper panel of their Fig. 5) and has remained steady with an 
average value of $1.25\pm 0.15$, indicating an essentially achromatic 
optical variability. 
However, a rather weak positive correlation between the color index and the 
flux was noticed in the study conducted by Gu et al.  (2006) covering a total 
of 50 nights (between Sept. 2003 - Feb. 2004), when source became bluer with 
increasing brightness (see, also, Hu et al. 2006).

\subsection{Long-term radio variability}

Since around 1979, PKS 0735+178 has been regularly monitored at 5, 8 and 15 GHz 
using the 26-metre Michigan dish. The observing technique, calibration 
procedures and instrumentation are described elsewhere (see, Aller et al. 
1985; 1999). The UMRAO light curves at 15 and 5 GHz are plotted in Figures 
1c \& 1d. Figure 1e shows the run of spectral index 
(defined as $S_{\nu}\propto{\nu^{a}}$) 
calculated by a linear regression analysis of the flux values at the three 
frequencies 
(Note that measurements at these frequencies were treated as simultaneous 
and hence combined only provided they are separated by no more than
$q = 0.04$ year). 
Figure 1f shows the variation of the percentage (linear) polarization at 15 GHz.
The monitoring of this blazar at still higher frequencies of 22 GHz and 37 GHz 
has been carried out using the 13.7-metre Mats\"ahovi antenna and the light 
curve for the period $\sim$ 1981 - 2004 is displayed in Fig. 1 of Ter\"asranta 
et al. (2005).  After a gradual decline from 1981 till 1987, the blazar 
displayed a large outburst when its flux at 37 GHz jumped from $\sim$1 Jy to 
$\sim$5 Jy. A second flare of comparable amplitude occured soon, following
which the flux declined to $\sim 1.5$ Jy by 1994.0. After the `twin radio 
flares', only mild variability has been observed. All these trends are 
closely mirrored in the UMRAO light curves at 15, 8 and 5 GHz (Fig. 1c \& d). 
From a comparison of the light curves at 5, 8, 15, 22 and 37 GHz, it is evident 
that the first of the twin radio flares has a flatter spectrum (i.e., a greater 
synchrotron opacity).
Interestingly, the pattern of the radio light curves is closely mimicked by 
the run of radio spectral index ($\alpha_{r}$).

Although a general increase in optical variability was noticed during the 
large radio outburst of 1987-1997 (Fig. 1b), no clear optical counterpart to
that radio outburst is evident from the data (Hanski et al.  2002; Tornikoski 
et al. 1994; Clements et al. 1995). In contrast, correlated flaring at optical 
and radio bands is known to be much more common for BL Lacs than for 
flat-spectrum radio quasars (FSRQs), suggesting that synchrotron opacity of 
the nuclear jet may be modest for the BL Lacs (Clements et al. 1995).

\subsection{Optical and radio polarization variability}

Characteristic of BL Lac objects, PKS 0735+178 has shown a large long-term 
variability of optical (linear) polarization, from about $1\%$ to more than
$30\%$ (Tommasi et al. 2001). However, on day-like or shorter time scales, 
covered in their polarimetric observations on four nights ( Dec. 10, 11, 
12 and 14, 1999), optical polarization remained steady at $\sim 10\%$, 
though a modest variability in the position angle was seen on internight and 
even intranight time scale. Thus, while the high polarization does signify
the active state of this blazar, the preferred polarization angle found over 
a few years (Tommasi et al. 2001) appears to manifest a state of stability 
in the optically emitting structures within the jet (Ciprini et al. 2007).
Note that a modest degree of optial polarization ($\sim 4\%)$ was 
measured in the observations of this LBL on Feb., 8, 2008 (Villforth et
al. 2009). 

The UMRAO time series of fractional (linear) polarization at 15 GHz, covering 
the period since 1981, is shown in Fig. 1f. Typically, the integrated 
polarization has remained between 2\% and 4\%, although near the first
peak of the twin radio flares in mid-1989, the 15 GHz polarization rose to 
5\% - 6\% level. Note that radio polarization at a few percent level with 
typical maximum value $< 10\%$, is a common occurrence for BL Lac objects 
(Aller et al. 1999). 
Correlation between the emergence of new VLBI component and the radio 
flaring has been reported by many authors (e.g., Wagner et al. 1995; Wehrle
, Pian \& Urry 1998) and usually these superluminal knots are substantially polarized 
(Gabuzda \& Cowthorne 1996; Lister, Marscher \& Gear 1998; G{\`o}mez, Marscher 
\& Alberdi 1999). Further, the study by Jorstad et al. (2001) has found 
that $ \gamma-$ray flares are correlated with the emergence of new VLBI 
components, which is usually explained in terms of the standard model where 
the VLBI components are the manifestations of a relativistic shock 
propagating through an underlying relativistic outflow (e.g., Marscher et al.
2008;  see, also, Krichbaum et al. 1995). 

\subsection{The variable compact radio jet}

Typical of blazars, the dominant radio core of PKS 0735+178 is surrounded by 
a $\sim 10''$ radio `halo' (Cassaro et al. 1999). Its (unbeamed) luminosity 
is at least an order-of-magnitude above the Fanaroff-Riley transition,
placing it in the Fanaroff-Riley class II (Fanaroff \& Riley 1974). 
A steep-spectrum radio jet of length $\sim 2''$ is seen to extend from the 
core at position angle (PA) $\sim 160^\circ$ (Tingay et al. 1998).
The core has been the target of many VLBI campaigns (e.g., Agudo et al. 
2006 and references therein). Multi-epoch VLBI imaging at 8, 22 and 43 GHz 
at several epochs from mid-1996 to mid-1998 revealed two peculiar sharp 
bends within the inner 2 milli arcsec ($\sim 11$ parsec) of the jet 
(G{\`o}mez et al. 1999; 2001; see, also, 
Kellermann et al. 1998). The consistency between these images and several 
previously reported, lower-resolution VLBI images taken during mid-1995, suggests
that the two bends were present already in mid-1995 (G{\`o}mez et al. 2001 and 
references therein). Intriguingly, this feature was absent in the VLBI images 
taken during the several years preceding mid-1992. The jet was then found 
to be rectilinear extending at PA  $\sim 65^\circ$ (B{\aa}{\aa}th \& Zhang 
1991; Gabuzda, Wardle \& Roberts 1989) and some of its knots exhibited 
large superluminal speeds ($7c - 12c$) (e.g., Gabuzda et al. 1994; G{\`o}mez 
et al. 2001 and references therein). Thus, sometime between mid-1992 and 
mid-1995, the blazar appears to have undergone a change into a regime 
characterized by quasi-stationary VLBI knots in the jet 
(G{\`o}mez et al. 2001; Agudo et al. 2006). Note that the ``transition epoch'' 
coincided with a huge decline in the radio flux that continued till 1998 
(Fig. 1c \& d). The sharp bends in the VLBI jet were probably caused by gas 
pressure gradients on 10-parsec scale within the nuclear region, since the 
magnetic field follows the jet curvature, thus supporting the notion of 
(non-ballistic) fluid motion along the jet during that
phase (G{\`o}mez et al. 1999).

Beginning with the year 2000, a resumption in the nuclear activity in this
blazar is manifested firstly by the upturn in its optical light curve (Fig. 
1a) and, secondly, by the near doubling of its 15 GHz flux density between 
mid-2000 and end-2002 (Fig. 1c \& d).  Interestingly, during the same 
period, the 15 GHz {\it MOJAVE} VLBI image (taken on 23 November 2002,
Lister \& Homan 2005) found  the jet to be fairly rectilinear and no longer 
exhibiting the double-bend witnessed during 1995 - 2000 within 3 milli 
arcsecond from core (e.g., see the 5 and 43 GHz VLBI images made by Agudo et 
al.  2006; also, Kellermann et al. 2004). The `straightened' VLBI jet with
PA $\sim 59-84^\circ$, as inferred from the all five components found within 
11 parsec from the core, is extremely mis-aligned from the kilo-parsec scale 
jet that extends at PA $\sim 105^\circ$ (Tingay et al. 1998). Large 
misalignment between the jets on parsec and kilo-parsec scales is indeed
quite common for LBL type BL Lac objects (Britzen et al. 2007 and references 
therein).

\section{Discussion}

In December 1998 when we embarked on the intranight optical monitoring of
PKS 0735+178, the $R$-band flux of this BL Lac had risen three-fold during 
the preceding year, from the historical minimum of 0.6 mJy to $\sim$ 2 mJy 
(Fig. 1a). 
By December 2001, the source had further brightened, forming a local peak of 
$\sim 6$ mJy which is only $\sim 3$ times lower than the historical maximum
attained by this blazar (Ciprini et al.  2007). Thus, the last decennium
has witnessed a ten-fold change in the optical (synchrotron) flux.  The 
factor would be even larger if the thermal accretion disk contributed
significantly near the brightness minimum.  In any event, the time span 
covered in our INOV monitoring witnessed a minimum three-fold rise in the 
optical synchrotron flux, followed by a similar amount of fading. Thus, in 
terms of its jet emission, the blazar cannot be deemed to have been quiescent 
during the decennium spanned by our monitoring programme. It is in this setting 
that a closer scrutiny of the INOV behavior of this blazar and of its
multi-band continuum and polarization properties is called for.

As seen from Table 1, PKS 0735+178 showed INOV on 4 out of the total 17 nights 
it was monitored in our programme. This corresponds to a INOV duty cycle DC = 
$27\%$ (see Sect. 3). Over these eleven years, the optical flux 
recorded a peak-to-peak amplitude variation of slightly over 1.7 mag (Fig. 3). 
Our data (Table 1) provide a hint that the nights of INOV detection coincide 
not with the extrema but with gradients in optical flux, consistent with the 
trend noted by Webb et al. (1998) and Howard et al. (2004). However, we 
caution that such a correlation is weak, at best, since the INOV detections 
were marginal.

The key result from our observations is that not even on one of the 17 nights 
was the INOV amplitude of this BL Lac found to be $> 3\%$, even though the 
probability of observing such INOV amplitudes is known to be $\sim 50\%$ in 
any single monitoring session longer than $\sim 4$ hours, a condition that was 
well satisfied in our programme (Sect. 1; Table 1). The probability of our 
negative result arising purely by chance is vanishingly small 
$< 7.6\times10^{-6}$.
This suggests that PKS 0735+178 has persisted in an INOV quiescent state
since 1998, despite other indications of its returning to an active state
during this period.
As summarized in Sect. 4, the indicators of renewed activity include (a) the 
large variation in its optical synchrotron flux on month/year-like time scale
and 
(b) the return of its VLBI jet to the `normal' rectilinear shape. Other 
indicators of a typical blazar state are the fairly high degrees of optical 
and radio polarization (Sect. 4.3) and the persistence of its flat radio 
spectrum (Fig. 1e). On the other hand, a rather uncharacteristic behaviour is
echoed by the fact that during the past two decades this blazar has undergone 
large radio outburst just once (Sect. 4.2; also, Hovatta et al. 2007) and
that  too without a clear optical counterpart (Sect. 4.3). Furthermore, its 
X-ray/$\gamma-$ray emission has been fairly steady (Sect. 1).

Here we note that recently from their 4-hour long $R$-band monitoring of this
blazar on Jan. 11, 2007 at Yunnan Observatory, Gupta et al. (2008) have 
reported an INOV detection. However, since their published DLC (Fig. 6 of 
their paper) is essentially flat and contains no significant structure on 
hour-like time scale, the INOV claimed by them can at best be of `flicker' 
type, with a time scale much shorter than the hour-like time scale we have
considered here.

In order to augment the present study we now turn attention to the (unpublished) 
optical monitoring data reported in Tables 3.1 and 4.1 of Dr. John Noble's
PhD dissertation (1995). During Jan. 25 - 31, 1992 he monitored PKS 0735+178 
on 6 nights in $R$-band, using the 1.07-metre Lowell telescope. On 5 of the 
nights, the monitoring duration was sufficiently long to meet our criterion
(6.7, 3.9, 4.2, 5.0 and 8.2 hours) and so also was the sensitivity of the
DLCs ($\sigma \leq 1\%$) (Sect. 1). Remarkably, on just one of these five 
nights did the blazar show INOV and that too with an amplitude of only $2\%$ 
(Table 4.1). This trend of INOV quiesence is further corroborated by Noble's 
5 hour long $V$-band monitoring of this blazar on March 17, 1989, using the KPNO 
0.9-metre telescope. Again, no INOV was detected, despite the high sensitivity 
attained ($\sigma = 1\%$). All these results place on a stronger footing the 
finding already emerging from our 17 nights' monitoring during 1998-2008 
(Table 1), namely that this {\it bona-fide} radio selected BL Lac is quite 
exceptional for its propensity to remain in a state of intranight optical 
quiescence (i.e., $\psi \leq 3\%$). Here it needs to be emphasized that, 
unlike our optical observations over the period 1998-2008, during which the
radio flux of PKS0735+178 was still at a comparatively low level (despite a 
resumption in its optical activity), the afore-mentioned optical data of
Noble (1995) are actually contemporaneous to the historically strongest
radio outburst (Fig.  1c \& d). 
Thus, while his March 1989 observations coincided with the first of the twin 
radio flares (the historical maximum), his 5 nights' monitoring during 
Jan. 1992 coincided with the onset of decline of the second peak of the twin 
radio flare (Fig. 1c \& d). 

{ \bf
Since substantial optical variability on wide ranging time scales,
which is typical of blazars, is commonly associated with shocks
forming and interacting with inhomogeneities in the Doppler boosted
synchrotron jets (e.g., Marscher, Gear \& Travis 1992), it is tempting 
to ask if the prolonged uncharacteristically subdued INOV level of
the classical blazar PKS 0735+178 is a manifestation of some unusual 
property of its jet. One model that explicitly attempts to address this 
question, particularly in the context of hour-like or shorter time 
scales of variability, invokes interaction of relativistic shocks in 
the jet with sub-parsec scale irregularities (Romero 1995). This scenario 
has been developed specifically for the two-fluid model of AGN jets 
originally proposed by Sol et al. (1989), wherein an extremely light 
and narrow beam of relativistic pair-plasma responsible for the apparent 
superluminal motion in the nuclear jets, is enveloped by a wider jet 
comprised of a much denser non-relativistic electron-proton plasma which 
carries most of the kinetic energy and terminates in kiloparsec sized 
hot spots (Pelletier \& Roland 1989; Henri \& Pelletier 1991). It has 
been argued that sub-parsec scale irregularities (needed for INOV) could 
arise in the beam of such jets due to classical macroscopic Kelvin-Helmholtz
instabilties, in case the axial magnetic field in the beam, $B_z$, is
below a critical value $B_c = 
[4\pi n_{b} m_{e} c^{2} (\gamma^2_{b}-1)]^{1/2}\gamma^{-1}_{b} $, where
$n_b$ and $\gamma_b$ are the electron number density and bulk Lorentz
factor of the beam, respectively and $m_e$ is the electron rest mass
(Romero 1995). Applying this model to the case of the well studied
intraday variable blazar 0917+624, for which estimates of the beam's
density and magnetic field are available, Romero has shown that for
a highly supersonic beam flow and equipartition magnetic fields, the 
(fastest growing) reflection modes would cross over into the unstable
regime and a rapid transition to a turbulent jet would occur, yielding 
the basic setting for INOV. Although estimates for the basic physical 
parameters for the present LBL PKS 0735+178 are not available, it is 
conceivable that its beam is fairly stable to the K-H instabilities 
(i.e., $B_z > B_c$, see above). Such a prospect is indeed supported
by the polarimetric VLBI data revealing that the magnetic field in the 
nuclear jet of this LBL is predominently axial (Agudo et al. 2001), 
unlike the norm for such BL Lacs (e.g., Lister \& Homan 2005 ; Kharb, 
Gabuzda \& Shastri 2008).
}

It is clear that in order to develop a proper understanding of the anomalous 
INOV behaviour of PKS 0735+178, clues will have to be gleaned 
by relating its multi-band variability patterns (of which some are distinctly 
typical of BL Lacs, while the others are less so, see above) to the VLBI
imaging and polarimetry of its nuclear jet. The VLBI observations have
already revealed enigmatic multiple twists in the nuclear jet which are
transitory, like the superluminal motion of its radio knots (Sect. 4).
Could the processes underlying this peculiar behaviour have kept
the inner jet (where INOV presumably originates) hidden from our view ?
These aspects will be examined by us elsewhere ({\bf Britzen et al. 2009, in prep.}), based on an available
sequence of VLBI images of this blazar taken at more than 20 epochs
over the past two decades.

\section*{Acknowledgments}
The authors are thankful to the anonymous referee for helpful
suggestions and to Dr. S. Ciprini for providing the $R$-band data. 
AG would like to thank Dr. Marcus Demleitner for providing the alpha 
version of Dexter's standalone version before it was released over ADS. 
M. Karouzos was supported for this research through a stipend from the 
International Max Planck Research School (IMPRS) for Astronomy and 
Astrophysics. UMRAO is funded by a series of grants from the NSF, most 
recently AST-0607523, and by funds from  the University of Michigan.
The 201-cm HCT is operated by the Indian Institute of Astrophysics (IIA).
The authors wish to acknowledge the support rendered by staff of IAO 
and CREST (IIA).

\clearpage

\begin{table*}
\caption{The observation log and INOV results for PKS 0735+178.}
\begin{tabular}{cccccccc}\\
\hline
 Date   & Telescopes used&   Number & Duration    & $\psi$  &  C$_{eff}$ & Status$^{\ddag}$ &  References\\
dd.mm.yy&          &  of points          &  (hours) & (\%)&            &                         \\
\hline
 26.12.98&    ST    & 49   &   7.8    &     1.8      &      1.13   &   N    &Sagar et al. (2004) \\ 
 30.12.99&    ST    & 64   &   7.4    &     1.0      &      0.61   &   N    &Sagar et al. (2004) \\ 
 25.12.00&    ST    & 42   &   6.0    &     1.6      &      1.02   &   N    &Sagar et al. (2004) \\ 
 25.12.01&    ST    & 43   &   7.3    &     1.0      &      2.8    &   V    &Sagar et al. (2004) \\ 
 20.12.03&   HCT    & 36   &   5.8    &     1.0      &      1.78   &   N    &present work     \\
 10.12.04&    ST    & 28   &   5.8    &     1.3      &      3.00   &   V    &present work     \\
 23.12.04&    ST    & 11   &   5.0    &     1.2      &      3.10   &   V    &present work     \\
 02.01.05&    ST    & 20   &   4.9    &     0.8      &      0.97   &   N    &present work     \\
 05.01.05&    ST    & 23   &   5.8    &     1.0      &      2.25   &  PV    &present work     \\
 09.01.05&    ST    & 28   &   6.7    &     1.3      &      3.20   &   V    &present work     \\
 09.11.05&    ST    & 17   &   3.8    &     0.7      &      2.00   &  PV    &present work     \\
 16.11.06&    ST    & 19   &   4.5    &     1.1      &      0.95   &   N    &present work     \\
 29.11.06&    ST    & 26   &   5.8    &     1.0      &      0.83   &   N    &present work     \\
 17.12.06&    ST    & 24   &   5.6    &     0.9      &      1.06   &   N    &present work     \\
 15.12.07&    ST    & 28   &   6.6    &     1.9      &      3.53   &   V    &present work     \\
 16.12.07&    ST    & 27   &   6.6    &     1.0      &      1.45   &   N    &present work     \\
 22.11.08&    ST    & 27   &   5.6    &     0.8      &      0.33   &   N    &present work     \\
\hline
\end{tabular}

{$^{\ddag}$V = variable; N = non-variable; PV = probable variable }
\end{table*}

\begin{table*}
\caption{Positions and apparent magnitudes of the comparison stars used in 
the present study (taken from United States Naval Observatory-B catalogue; Monet et al. 2003)}
\begin{tabular}{cccccc}\\
\hline
Star & RA(J2000) &Dec(J2000) &  {\it B} & {\it R} & \it {B-R} \\
     &           &            & (mag)    & (mag)   & (mag) \\
\hline
S1   & 07{$^h$}37{$^m$}48{$^s$}.21 & $+$17{$^\circ$}40{$^\prime$}14{$^{\prime\prime}$}.4 & 16.05 & 14.86& 1.19\\
S2   & 07{$^h$}38{$^m$}12{$^s$}.15 & $+$17{$^\circ$}38{$^\prime$}07{$^{\prime\prime}$}.3 & 15.62 & 15.25& 0.37\\
S3   & 07{$^h$}37{$^m$}48{$^s$}.88 & $+$17{$^\circ$}41{$^\prime$}30{$^{\prime\prime}$}.4 & 16.23 & 15.26& 0.97\\
S4   & 07{$^h$}38{$^m$}26{$^s$}.00 & $+$17{$^\circ$}40{$^\prime$}22{$^{\prime\prime}$}.5 & 16.15 & 15.55& 0.60\\
S5   & 07{$^h$}38{$^m$}13{$^s$}.82 & $+$17{$^\circ$}38{$^\prime$}22{$^{\prime\prime}$}.8 & 15.29 & 14.68& 0.61\\
S6   & 07{$^h$}38{$^m$}08{$^s$}.35 & $+$17{$^\circ$}44{$^\prime$}59{$^{\prime\prime}$}.3 & 15.94 & 15.59& 0.35\\
SS1  & 07{$^h$}38{$^m$}03{$^s$}.42 & $+$17{$^\circ$}42{$^\prime$}55{$^{\prime\prime}$}.8 & 16.48 & 15.89& 0.59\\
SS2  & 07{$^h$}38{$^m$}17{$^s$}.08 & $+$17{$^\circ$}39{$^\prime$}03{$^{\prime\prime}$}.6 & 16.29 & 15.96& 0.33\\
SS3  & 07{$^h$}38{$^m$}10{$^s$}.26 & $+$17{$^\circ$}43{$^\prime$}43{$^{\prime\prime}$}.8 & 16.71 & 16.46& 0.25\\
\hline
\end{tabular}
\end{table*}

\clearpage
\newpage
\begin{figure*}
\includegraphics[height=23.0cm,width=18.0cm]{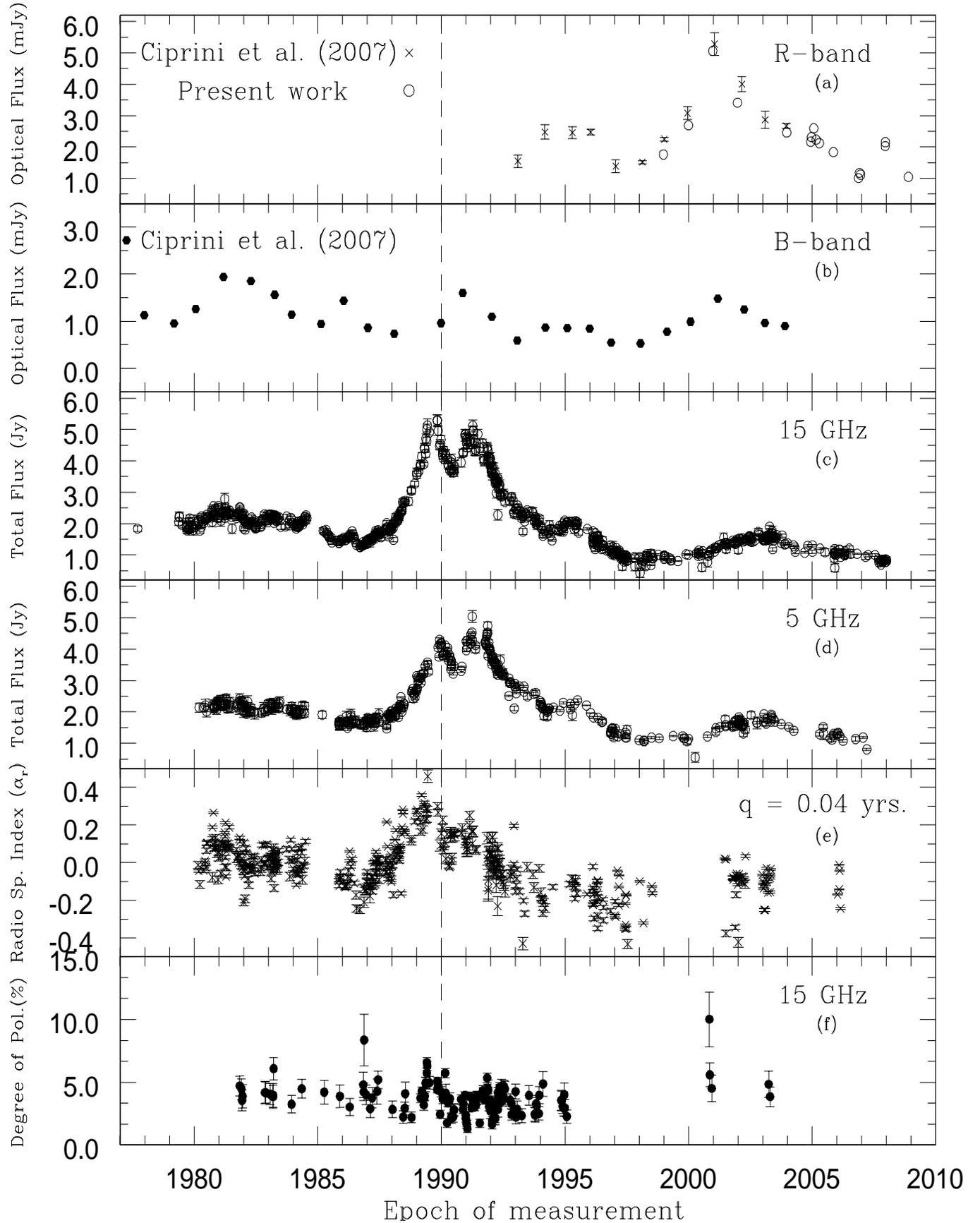}
\caption { Light curves of PKS 0735+178 (Sect. 4); the vertical broken line 
marks the epoch of its historical maximum in radio brightness.}
\end{figure*}
\clearpage
\newpage
\begin{figure*}
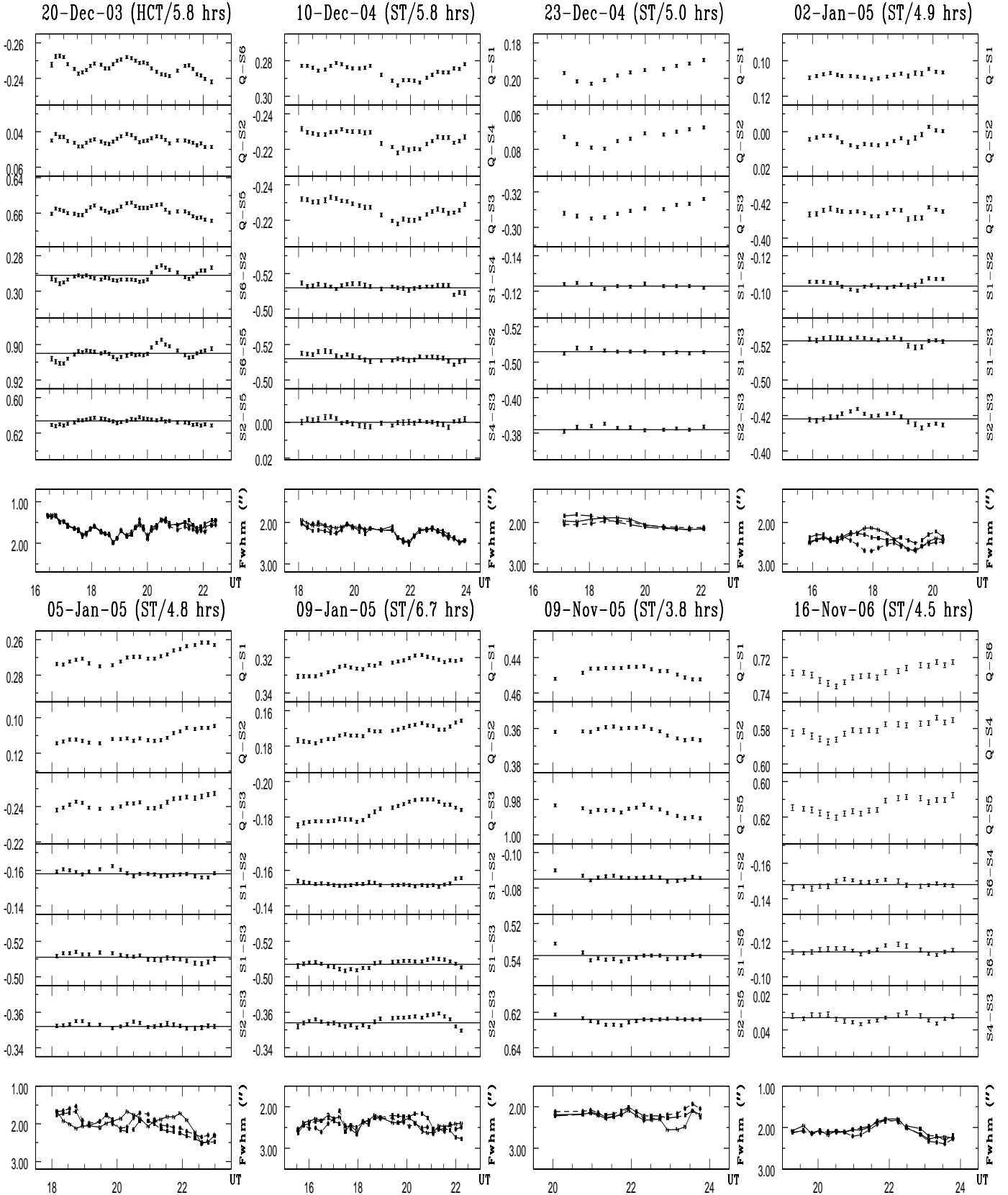

\hspace*{-1.0cm}
\hbox{
\includegraphics[height=11.0cm,width=04.5cm]{fig_0735+178_HCT_20dec03.epsi}
\includegraphics[height=11.0cm,width=04.5cm]{fig_0735+178_ST_10dec04.epsi}
\includegraphics[height=11.0cm,width=04.5cm]{fig_0735+178_ST_23dec04.epsi}
\includegraphics[height=11.0cm,width=04.5cm]{fig_0735+178_ST_02jan05.epsi}
}
\vspace*{1cm}
\hspace*{-1.0cm}
\hbox{
\includegraphics[height=11.0cm,width=04.5cm]{fig_0735+178_ST_05jan05.epsi}
\includegraphics[height=11.0cm,width=04.5cm]{fig_0735+178_ST_09jan05.epsi}
\includegraphics[height=11.0cm,width=04.5cm]{fig_0735+178_ST_09nov05.epsi}
\includegraphics[height=11.0cm,width=04.5cm]{fig_0735+178_ST_16nov06.epsi}
}
\caption{The intranight DLCs of PKS 0735+178. For each night the upper three 
panels show the DLCs of the blazar relative to three steady comparison stars 
while the lower three panels show the star-star DLCs. The bottom panel gives 
the plots of seeing variation for the night, based on three stars. For each 
night, the date, duration of monitoring and the telescope used are mentioned
near the top.}
\label{fig:1}
\end{figure*}
\clearpage
\newpage
\begin{figure*}
\hbox{
\hspace*{-1.0cm}
\includegraphics[height=11.0cm,width=04.5cm]{fig_0735+178_ST_29nov06.epsi}
\includegraphics[height=11.0cm,width=04.5cm]{fig_0735+178_ST_17dec06.epsi}
\includegraphics[height=11.0cm,width=04.5cm]{fig_0735+178_ST_15dec07.epsi}
\includegraphics[height=11.0cm,width=04.5cm]{fig_0735+178_ST_16dec07.epsi}
}
\vspace*{1cm}
\hspace*{-1.0cm}
\hbox{
\includegraphics[height=11.0cm,width=04.5cm]{fig_0735+178_ST_22nov08.epsi}
}
\begin{center}
{{\bf Figure~\ref{fig:1}}. \textit {continued}}
\end{center}
\end{figure*}

\clearpage
\newpage
\begin{figure*}
\hspace*{-0.5cm}
\includegraphics[height=18.0cm,width=18.0cm]{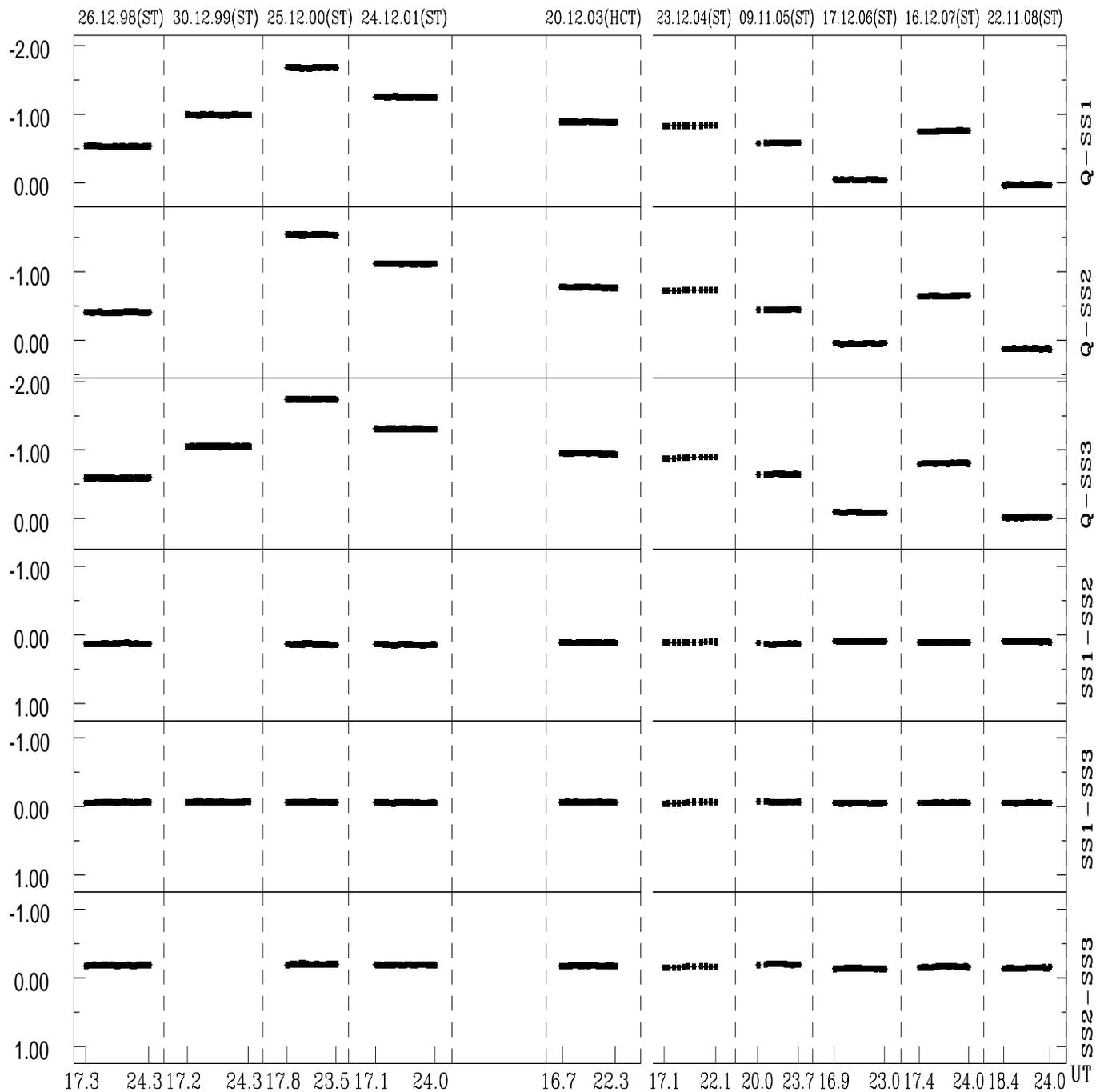}
\caption {The long-term $R$-band differential light curves of PKS 0735+178,
derived from from our data spanning 11 years (1998-2008).  In order to maintain
consistency with our already published data (Sagar et al. 2004), we have used 
the same three comparison stars for generating these LTOV plots.
Each horizontal sedgement of the light curves represents the DLC
for the night corresponding to the date mentioned at the top, along
with the telescope used (shown in parentheses).
} 
\end{figure*}

\clearpage
\newpage
\begin{figure*}
\hspace*{-0.5cm}
\includegraphics[height=15.0cm,width=15.0cm]{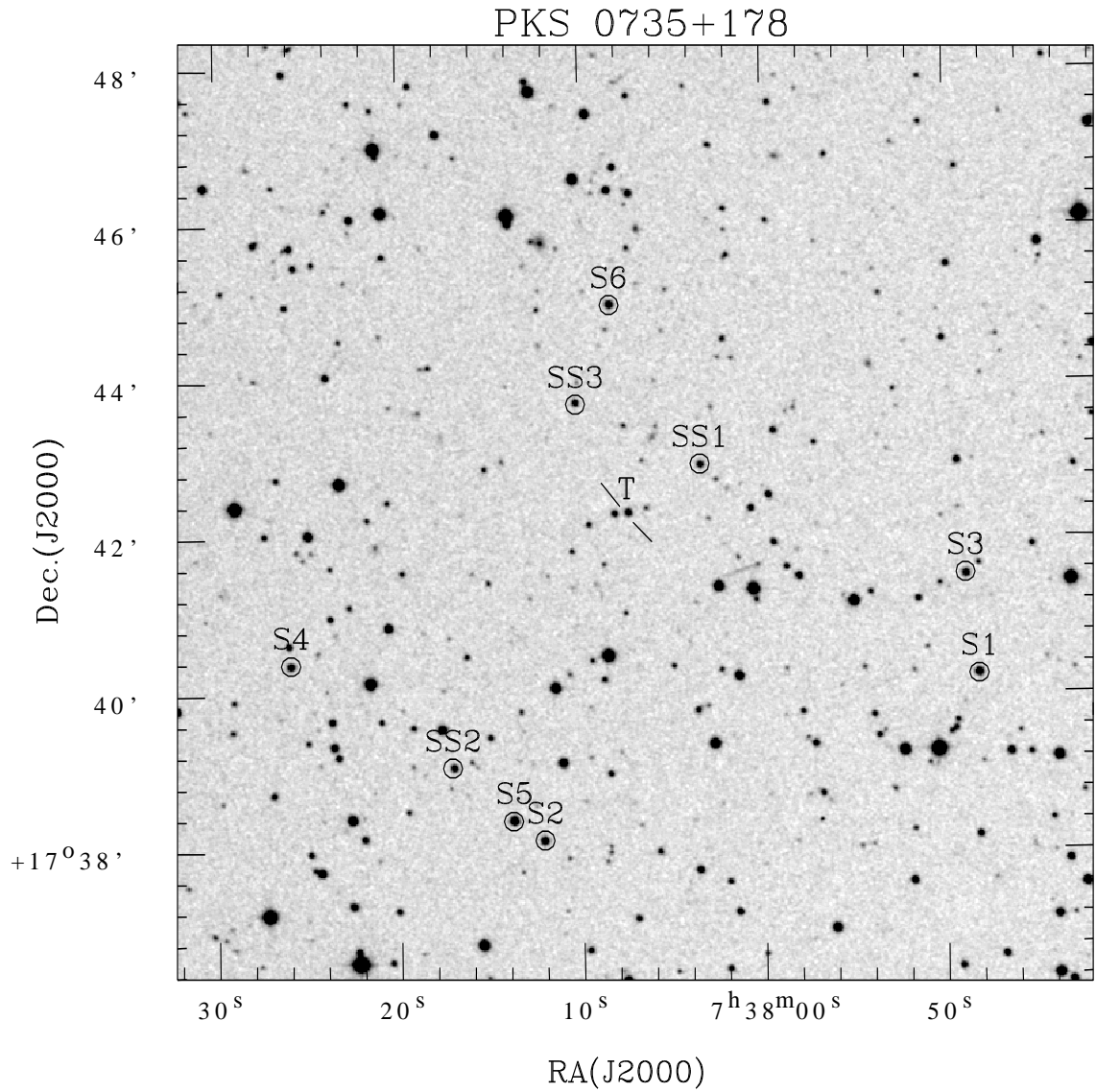}
\caption {
DSS POSS2 $R$-band $12^\prime \times 12^\prime$ field centered on PKS 0735+178 is shown.
The positions of target AGN (marked with T, within double bar) and
the comparison stars is shown within circles and marked with S1...S6 and SS1..SS3
notations used for making INOV \& LTOV plots, respectively.
}

\end{figure*}
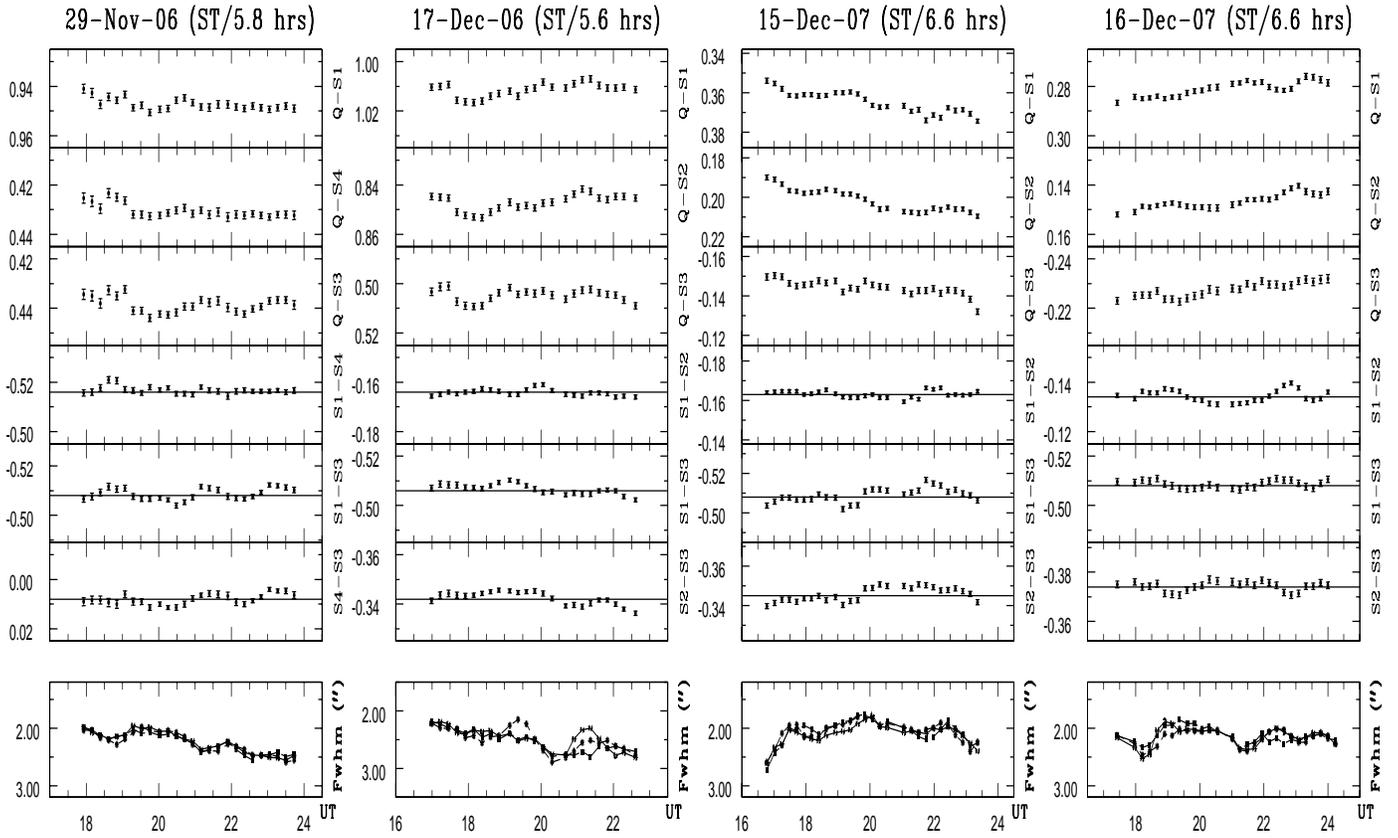
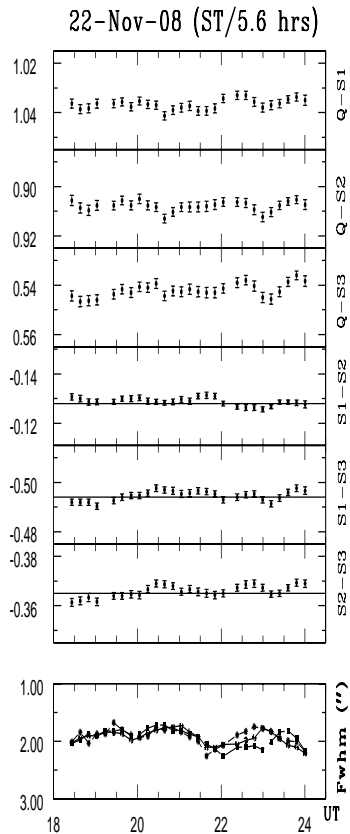
\end{document}